\documentclass[11pt,a4paper]{article}
\usepackage{graphicx}
\usepackage{amsmath}
\usepackage{amssymb}
\usepackage{cite}
\usepackage{subcaption}

\newcommand{\bm}{\boldsymbol}
\newcommand{\vect}[1]{\boldsymbol{#1}}
\newcommand{\mat}[1]{\boldsymbol{#1}}

\newcommand{\Proj}[1]{\mat{P}_{#1}}

\newcommand{\rank}[1]{\mathrm{rank}\left(#1\right)}
\newcommand{\range}[1]{\mathcal{R}\left\{#1\right\}}

\newcommand{\sqnorm}[1]{\left\|#1\right\|^{2}}

\newcommand{\mOmega}{\bm{\Omega}}
\newcommand{\mPsi}{\bm{\Psi}}
\newcommand{\mSigma}{\bm{\Sigma}}

\newcommand{\vd}{\vect{d}}
\newcommand{\vq}{\vect{q}}
\newcommand{\vv}{\vect{v}}
\newcommand{\vw}{\vect{w}}
\newcommand{\vwtilde}{\tilde{\vw}}
\newcommand{\vwtildea}{\vwtilde_{a}}
\newcommand{\wtildemn}{\vwtilde_{\text{\tiny{MN}}}}

\newcommand{\A}{\mat{A}}
\newcommand{\B}{\mat{B}}
\newcommand{\C}{\mat{C}}
\newcommand{\G}{\mat{G}}
\newcommand{\I}{\mat{I}}
\newcommand{\eye}[1]{\I_{#1}}
\newcommand{\Q}{\mat{Q}}
\newcommand{\V}{\mat{V}}
\newcommand{\Vorth}{\V_{\perp}}
\newcommand{\X}{\mat{X}}
\newcommand{\Xt}{\X_{t}}
\newcommand{\Z}{\mat{Z}}
\newcommand{\Ztilde}{\tilde{\Z}}

\newcommand{\pdfN}[4]{\mathcal{N}_{#1}\left(#2,#3,#4\right)}
\newcommand{\dist}{\overset{d}{=}}

\begin{document}
\title{Partially adaptive filtering using randomized projections}
\author{Olivier Besson}
\date{Universit\'{e} de Toulouse, ISAE-SUPAERO }
\maketitle
\begin{abstract}
This short note addresses the design of a partially adaptive filter to retrieve a signal of interest in the presence of strong low-rank interference and thermal noise. We consider a generalized sidelobe canceler implementation where the dimension-reducing transformation is build resorting to ideas borrowed from randomized matrix approximations. More precisely, the main subspace of the auxiliary data $\Z$ is approximated by $\Z\mOmega$ where $\mOmega$ is a random matrix or a matrix that picks at random columns of $\Z$.  These transformations do not require eigenvalue decomposition, yet they provide performance similar to those of a principal component filter. 
\end{abstract}


\section{Problem statement}
\begin{figure}[h]
\centering
\includegraphics[width=8cm]{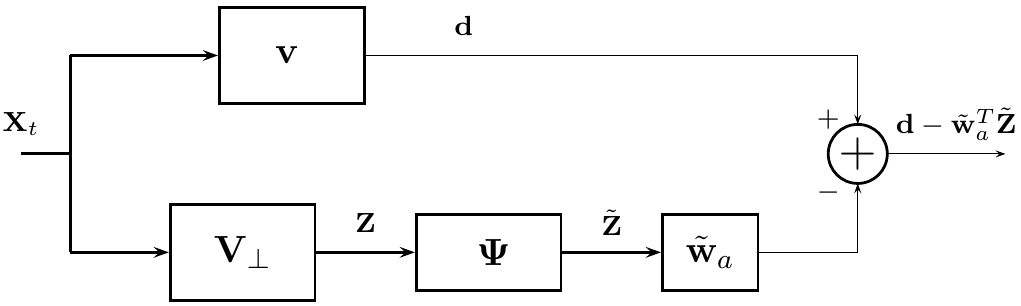} 
\caption{Structure of a partially adaptive filter with $\mPsi$ a $(N-1) \times R$ matrix.}
\label{fig:structure_pa_df}
\end{figure}

In many fields of engineering, including radar, communications, hyperspectral imaging, it is desired to extract as best as possible a known signal of interest (SoI) among disturbance (interference plus thermal noise) with unknown statistics. A simple and yet very effective way to achieve this goal consists in designing a linear adaptive filter which at the same time preserves the SoI and strongly attenuates  interference. As predicted by the Reed Mallett and Brennan rule \cite{Reed74}, approximately $2N$ samples -with $N$ being the size of the observations- are required for an adaptive filter to provide the optimal signal to noise ratio (SNR) up to $3$dB. When $N$ is large or training samples are scarce this requirement may not be fulfilled. This is the situation we consider herein where the number of training samples available $K < N$. A possible solution lies in a partially adaptive filter \cite{Ward94,Goldstein97c} which relies on reducing the size of the observations -by means of a linear transformation- and then operating in this lower dimensional space. This approach is illustrated in Figure \ref{fig:structure_pa_df}  with a generalized sidelobe canceler structure. 
Here $\vv$ is the signature of the SoI (assumed to be unit-norm) and $\Vorth$ is a semi-unitary matrix orthogonal to $\vv$. $\Xt$ stands for the $N \times K$ matrix of the training samples and $\Z = \Vorth^{T}\Xt$ corresponds to the auxiliary channels.  $\mPsi$ denotes the $(N-1) \times R$ matrix operating on $\Z$ which achieves dimension reduction. Its goal is to capture the interference subspace so that the reduced-dimension adaptive filter $\vwtildea$ can estimate from $\Ztilde=\mPsi^{T}\Z$ the part of interference present in the main channel $\vd=\vv^{T}\Xt$. 

This structure is especially efficient when the disturbance covariance matrix $\mSigma$ is the sum of a low rank term plus a scaled identity matrix i.e., $\mSigma=\G\G^{T}+\sigma^{2}\eye{N}$ where $\G$ is a $N \times J$ matrix. Actually if $\mSigma$ is known the optimal filter $(\vv^{T}\mSigma^{-1}\vv)^{-1}\mSigma^{-1}\vv$ coincides with a partially adaptive filter where $\mPsi = \Vorth^{T}\G$ \cite{Ward94}. In practical situations where $\mSigma$ is unknown, this partially adaptive filter can result in nearly optimal performance, provided of course that $\mPsi$ captures most of the range space of $\Vorth^{T}\G$. Actually, the SNR loss at the output of a partially adaptive filter with a fixed $\mPsi$ is distributed according to a scaled beta distributed random variable where the scaling factor $a$ measures how much of the main interference subspace is retained through $\mPsi$ \cite{Bessonsnrloss}. As illustrated in \cite{Bessonsnrloss} the value of $a$ is close to $1$ whenever the angles between $\range{\Vorth^{T}\G}$ and $\range{\mPsi}$ are not too large: in \cite{Bessonsnrloss} it was shown that $a \geq 0.95$ when these angles are distributed over $[0^{\circ},45^{\circ}]$. In other words, the matrix $\mPsi$ needs to retain most of the main subspace of $\Z$ without the need for high accuracy at this step since there is a sort of post-processing through $\vwtildea$ that can accommodate small differences. 

Since the goal of $\mPsi$ is that most of the main subspace of $\Z$ carries on to $\Ztilde$, a logical choice is to select the $R$ principal left singular vectors of $\Z$ as the columns of $\mPsi$. This choice leads to the so-called principal component method \cite{Kirsteins91,Haimovich96,Haimovich97}.  An alternative choice is to choose the left singular vectors which result in highest SNR, as in the cross spectral metric method \cite{Goldstein97b}. Now these methods require singular value decomposition (SVD) which can be computationally demanding. In the sequel we investigate an approach based on randomized methods which does not require SVD but provides a sufficiently accurate estimate of $\range{\Vorth^{T}\G}$ for the purpose of partially adaptive filtering.

\section{Design of $\mPsi$ using randomized projections}
In what follows we introduce some randomness in the selection of $\mPsi$. The idea of using a random matrix $\mPsi$, drawn from a complex Gaussian distribution, has been proposed and analyzed by Marzetta \cite{Marzetta11}. However when using a completely random $\mPsi$  then it is rather unlikely that $\mPsi$ will retain most of the energy in the principal subspace of $\Z$. This is why Marzetta proposed to use many matrices $\mPsi_{\ell}$ and to average the corresponding partially adaptive filters. Herein, we use a single matrix $\mPsi$ but it is constructed from $\Z$ so as to guarantee the desired property. 

In recent years there has been a growing interest in random methods to provide low-rank approximations of matrices \cite{Halko11,Martinsson19,Martinsson21}.  More precisely, for a given $m \times n$ matrix, one wishes to find an approximation $\A \approx \underset{m \times k}{\B} \; \underset{k \times n}{\C}$. The approximation is usually computed in two stages. A first stage consists of a \emph{rangefinder}, whose goal is to construct a low-dimensional subspace that captures most of $\range{\A}$. In other words one wants to identify a $m \times (k+p)$ matrix $\Q$ with orthogonal columns such that $\A \approx \Q \Q^{T}\A$.  Towards this end $\Q$ is generally obtained from a QR decomposition of $\A\mOmega$ where $\mOmega$ is a $n \times (k+p)$ random matrix. The second stage  consists of reduced SVD computation of $\Q^{T}\A$ \cite{Halko11}. 

For the purpose of partially adaptive filtering, only the first step is necessary since we want $\mPsi$ to retain the principal subspace of $\Z$. Moreover, in contrast to the randomized low-rank approximation, we do not need an approximate orthogonal basis for the subspace to be approximated. This suggests to use
\begin{equation}\label{Psi=Z_Omega}
\mPsi = \Z \mOmega
\end{equation}
where $\mOmega$ is a $K \times R$ random matrix. If the covariance matrix of $\Xt$ is $\mSigma=\G\G^{T}+\sigma^{2}\eye{N}$ with $\rank{\G}=J$ then it can be surmised that most of the subspace where $\Z$ lies will be retained provided that $R \geq J$. With this choice, one has $\Ztilde = \mPsi^{T}\Z = \mOmega^{T}\Z^{T}\Z$. The vector $\vwtildea$, which is obtained by minimizing 
\begin{equation}\label{prob_pa}
\sqnorm{\Ztilde^{T}\vwtildea - \vd^{T}}  = \sqnorm{\Z^{T}\mPsi\vwtildea - \vd^{T}}
\end{equation}
can thus be written as
\begin{align*}
\vwtildea &= (\Ztilde\Ztilde^{T})^{-1}\Ztilde \vd^{T} \nonumber \\
&= 	(\mOmega^{T}\Z^{T}\Z\Z^{T}\Z \mOmega) ^{-1} \mOmega^{T}\Z^{T}\Z \vd^{T}
\end{align*}
The equivalent length-$N$ filter is then given by
\begin{align}
\vw &= \vv - \Vorth \mPsi \vwtildea \nonumber \\
&= \vv - \Vorth \Z \mOmega (\mOmega^{T}\Z^{T}\Z\Z^{T}\Z \mOmega) ^{-1} \mOmega^{T}\Z^{T}\Z \vd^{T} \nonumber \\
&= \vv - \Vorth \Z (\Z^{T}\Z) ^{-1} \Proj{\Z^{T}\Z\mOmega} \vd^{T}
\end{align}
As for the matrix $\mOmega$ it can be possibly drawn from a Gaussian distribution, i.e., $\mOmega$ has independent entries drawn from a Gaussian distribution with zero mean and unit variance. Another possibility we will explore is to set $\mOmega(i_{r},r)=1$ and $0$ otherwise, where $(i_{1},\ldots,i_{R})$ is a random partition of $(1,\ldots,K)$. This idea amounts to select only $R$ columns of $\Z$, i.e., $\mPsi = \begin{bmatrix} z_{i_1} & z_{i_2}& \cdots & z_{i_R} \end{bmatrix}$. We will investigate in the next section the performance of each of these two choices.

As a final comment, we note that the case of interest here is when $K$ is small, typically of the order of $J-2J$. As suggested in \cite{Halko11,Martinsson19,Martinsson21} it is relevant to choose $R$ slightly above $J$ which means that $R$ could potentially be close to $K$. In the limiting case $R=K$,  $\mOmega$ is irrelevant and $\vw$ does not longer depend on $\mOmega$. Actually it becomes
\begin{equation}\label{w_MN}
\vw_{(K=R)}  = 	 \vv - \Vorth \Z \ (\Z^{T}\Z) ^{-1}  \vd^{T} =  \vv - \Vorth \wtildemn
\end{equation}
where
\begin{equation}
\wtildemn =  \arg \underset{\Z^{T}\vwtilde = \vd^{T}}{\min} \sqnorm{\vwtilde} 
\end{equation}
is the minimum norm vector which satisfies $\Z^{T}\vwtilde = \vd^{T}$. In the sequel we will study the performance of this method. 
 \section{Simulations}
We now evaluate the performance of the proposed method and compare them with those of the principal component (PC) filter.  We consider a scenario with $N=100$. The disturbance covariance matrix is given by $\mSigma = \sum_{j=1}^{J}\lambda_{j}\vq_{j}\vq_{j}^{T} + \I$ where $10\log_{10}\lambda_{j}$ is drawn from a uniform distribution over $[15,25]$dB. The matrix $\Q=\begin{bmatrix} \vq_{1} & \vq_{2} & \cdots & \vq_{J} \end{bmatrix}$ of eigenvectors is drawn randomly on the Stiefel manifold. The angle between $\vv$ and $\Q$ is denoted $\theta$, and is set to $\theta=75^{\circ}$.

\begin{figure}[h]
\centering
\includegraphics[width=8cm]{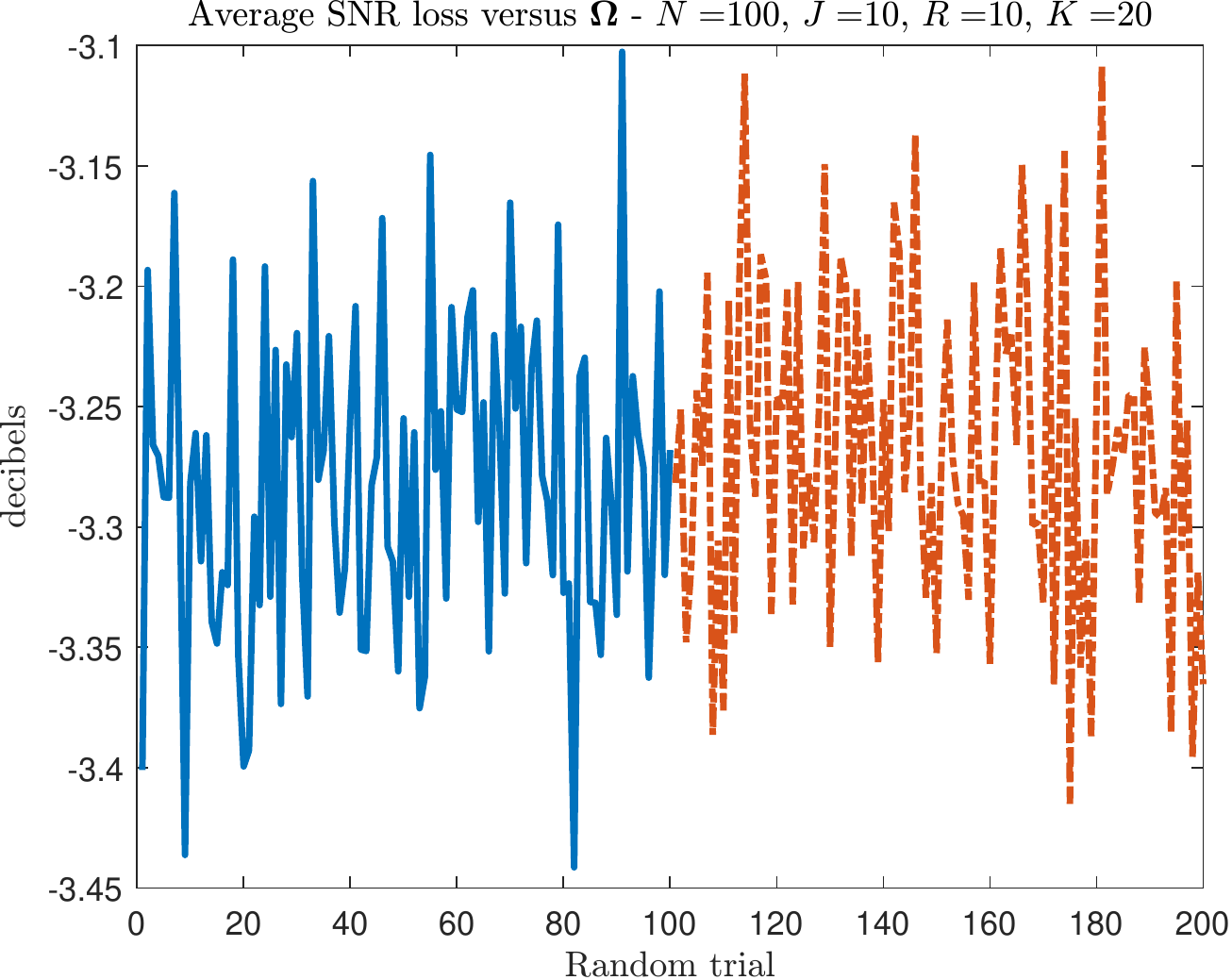}	
\caption{Average SNR loss versus $\mOmega$.  $\mPsi=\Z\mOmega$ with $\mOmega \dist \pdfN{K,R}{\mat{0}}{\I_{K}}{\I_{R}}$ in the first $100$ trials,$\mPsi = \begin{bmatrix} z_{i_1} & z_{i_2}& \cdots & z_{i_R} \end{bmatrix}$ in the $100$ last trials. $N=100$, $J=10$, $R=J$ and $K=2J$.}
\label{fig:mean_snrloss_vs_Omega_theta=75_R=J_K=2}
\end{figure}

First we investigate the influence of $\mOmega$ on the performance of the proposed methods. In Figure \ref{fig:mean_snrloss_vs_Omega_theta=75_R=J_K=2} we display the average SNR loss for $200$ different matrices $\mOmega$. In the first $100$ trials $\mOmega$ is drawn from a Gaussian distribution while in the last $100$ trials $\mPsi$ consists of $R$ columns of $\Z$ drawn randomly. As can be observed, the average SNR loss depends rather weakly on $\mOmega$ and we do not see any significant difference between the two choices. This indicates that \emph{the performance is nearly independent of $\mOmega$}, which is an appealing feature. 

\begin{figure}[htb]
	\centering
	\begin{subfigure}[b]{0.48\textwidth}
		\centering
		\includegraphics[width=\textwidth]{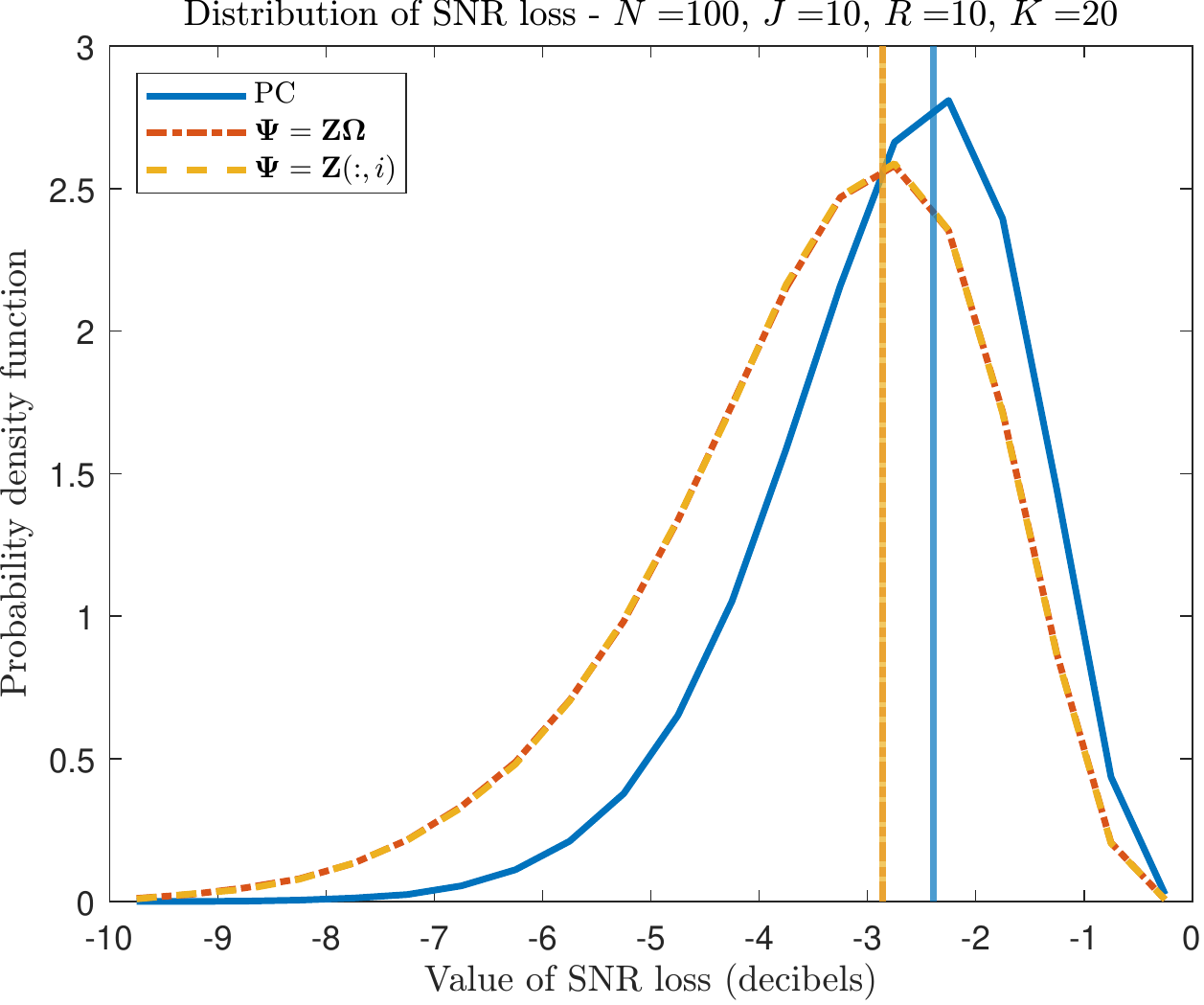}
		\caption{$J=10$, $R=10$}
	\end{subfigure}
	\hfill
	\begin{subfigure}[b]{0.48\textwidth}
		\centering
		\includegraphics[width=\textwidth]{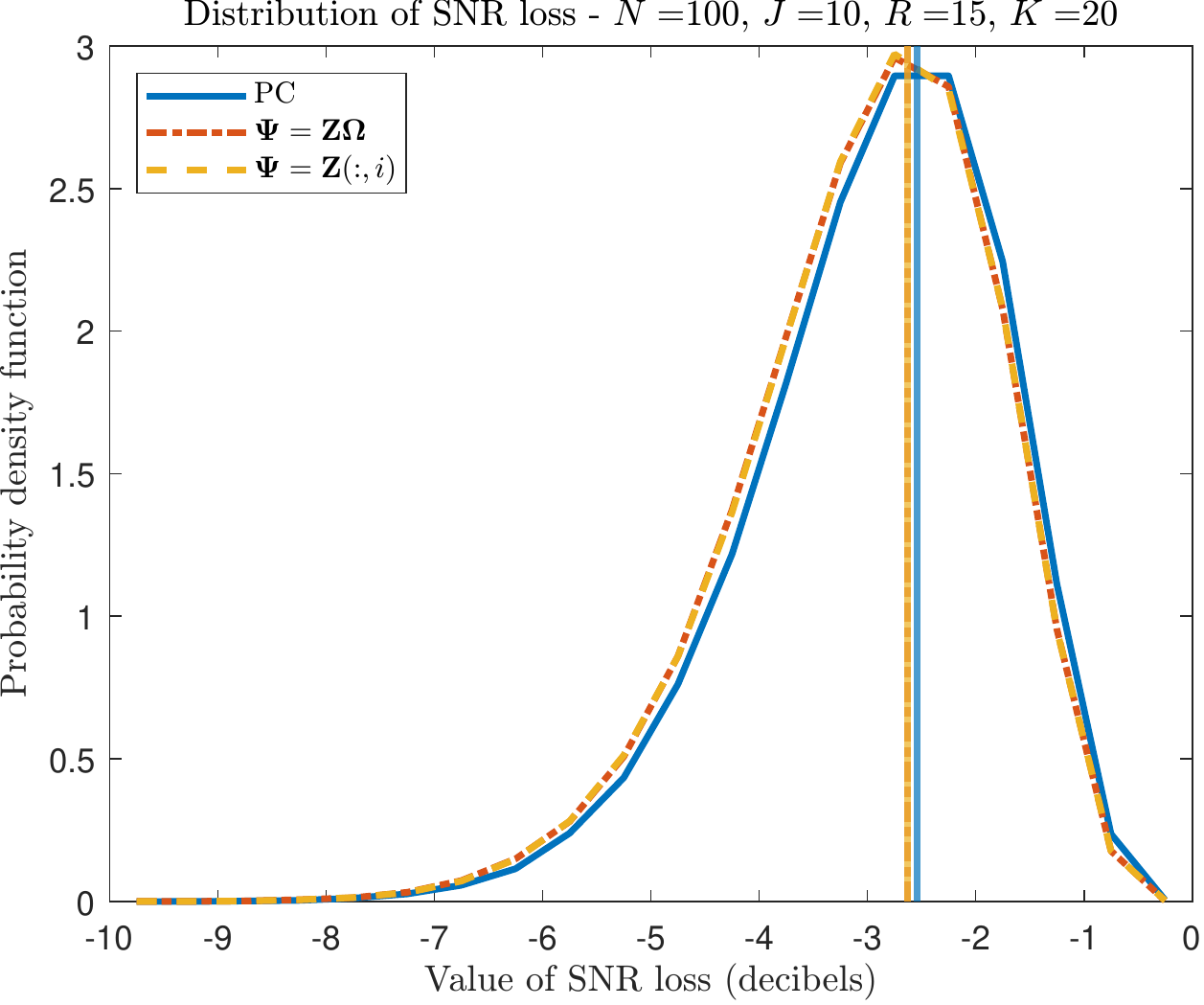}
		\caption{$J=10$, $R=15$}
	\end{subfigure}
	\\
	\begin{subfigure}[b]{0.48\textwidth}
		\centering
		\includegraphics[width=\textwidth]{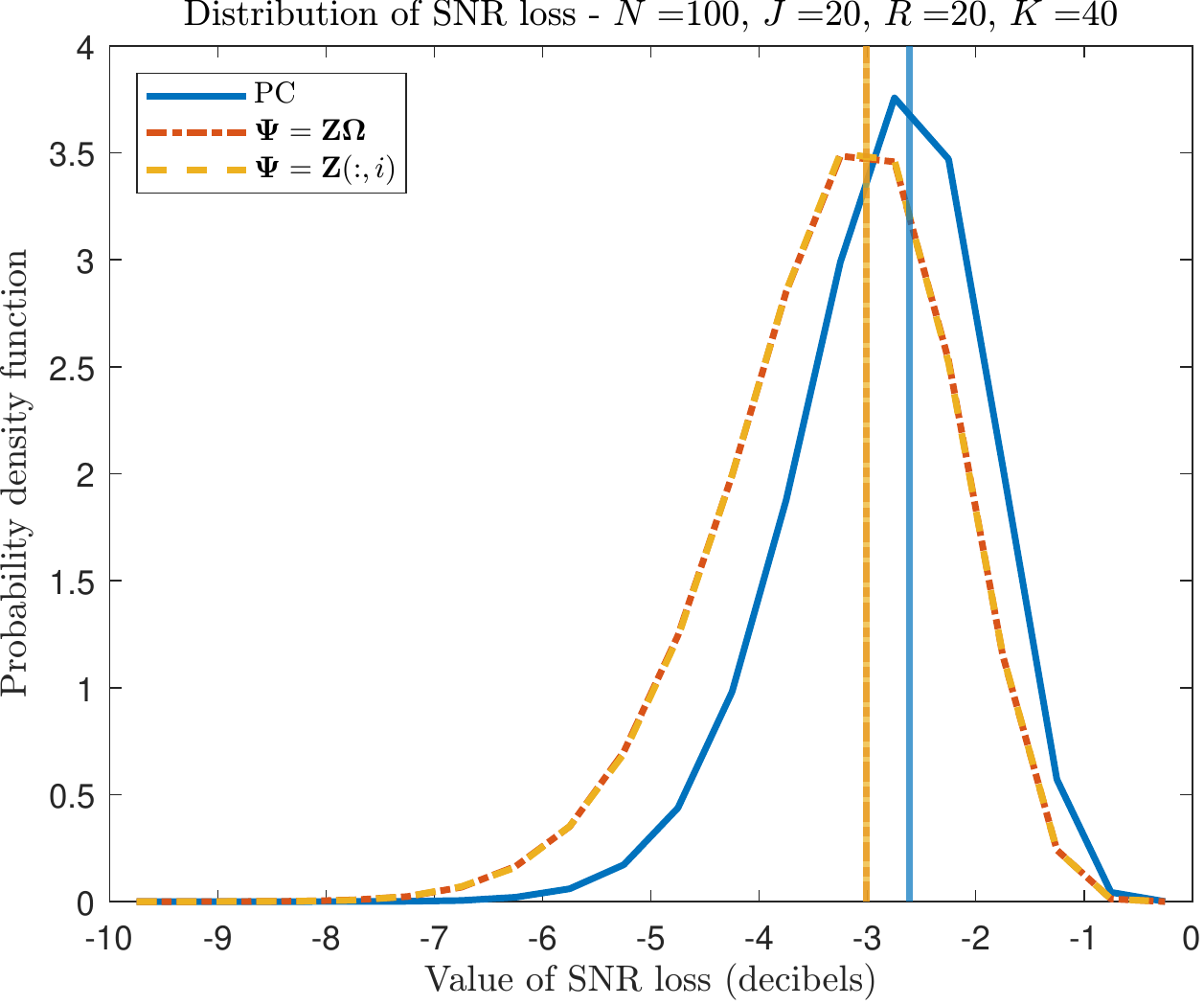}
		\caption{$J=20$, $R=20$}
	\end{subfigure}
	\hfill
	\begin{subfigure}[b]{0.48\textwidth}
		\centering
		\includegraphics[width=\textwidth]{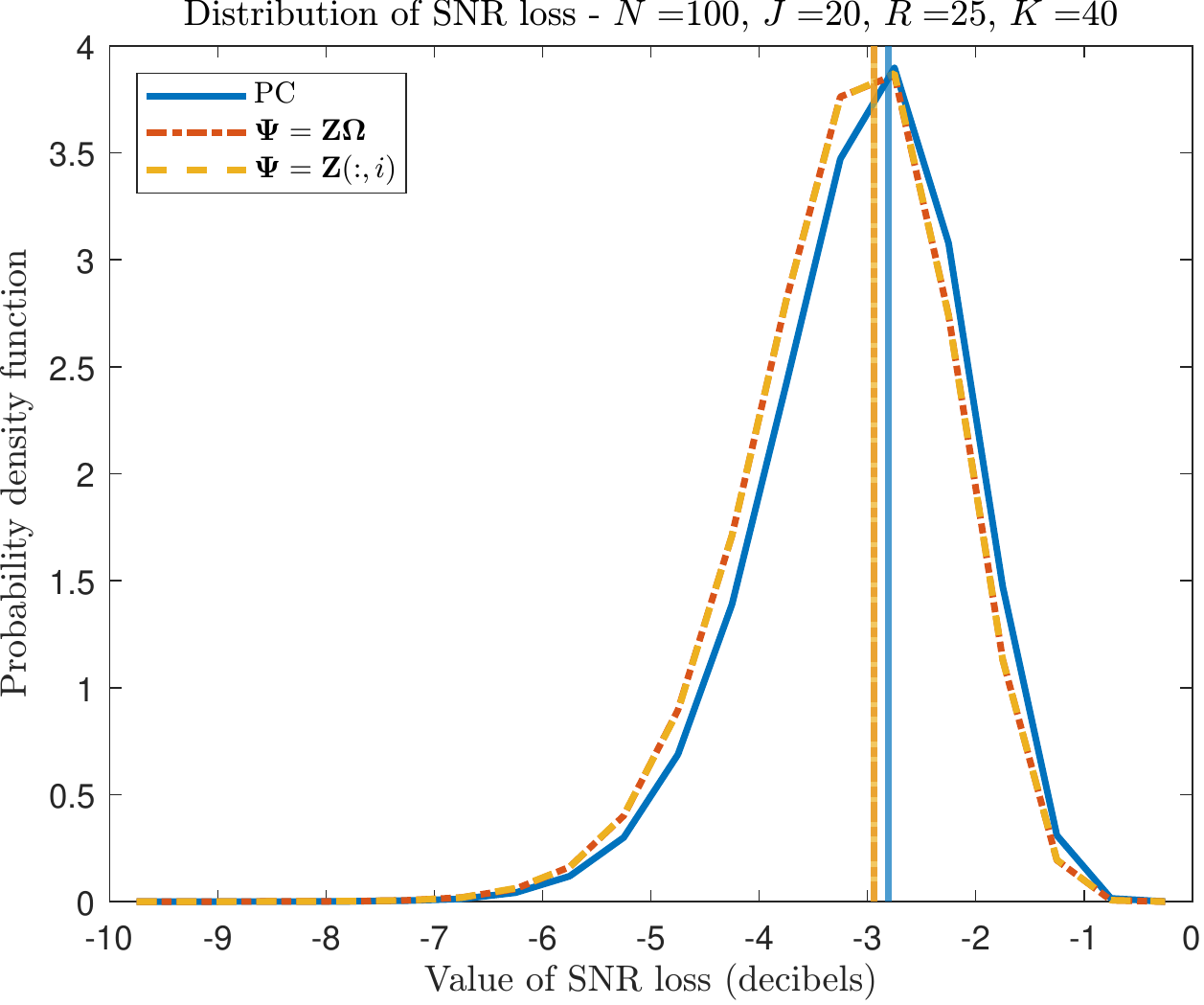}
		\caption{$J=20$, $R=25$}
	\end{subfigure}
	\caption{Distribution of the SNR loss. $N=100$ and $K=2J$.}
	\label{fig:pdf_snrloss}	
\end{figure}

In Figure \ref{fig:pdf_snrloss} we display the distribution of the SNR loss for the proposed methods based on $\mPsi=\Z\mOmega$ and  the PC method. The vertical lines represent the average value of the SNR loss. As can be seen from this figure, if $R$ is chosen as $R=J$, the methods based on $\mPsi=\Z\mOmega$ come closer to the PC method when $J/N$ increases. An interesting observation is also that it is beneficial to use $R > J$, which agrees with what is actually recommended in random methods for low-rank matrix approximation \cite{Halko11,Martinsson19,Martinsson21}.  Finally, we do not notice any difference between the two choices of $\mOmega$.

\begin{figure}[htb]
\centering
\begin{subfigure}[b]{0.48\textwidth}
\centering
\includegraphics[width=\textwidth]{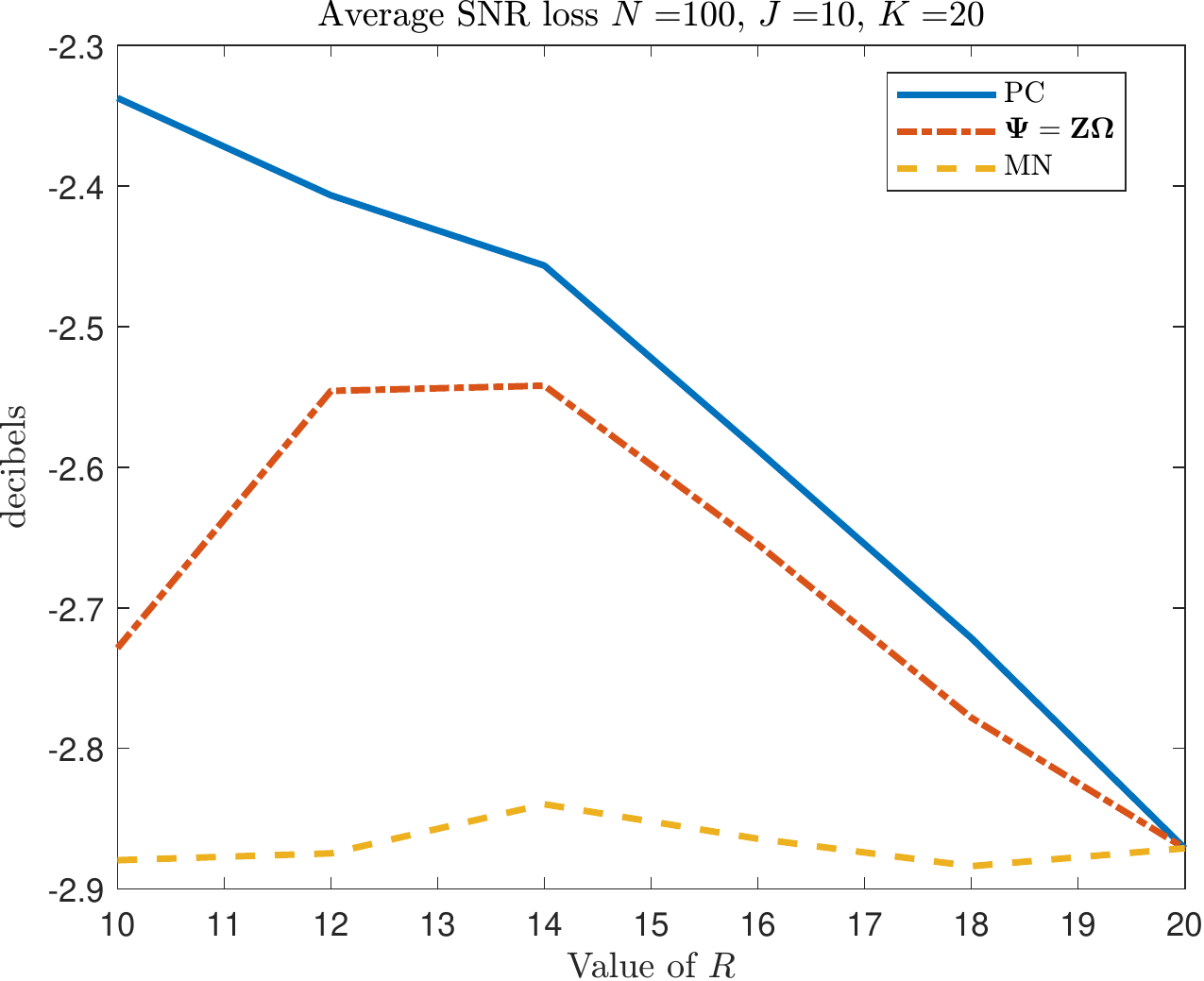}
\caption{$J=10$, $K=20$}
\end{subfigure}
\hfill
\begin{subfigure}[b]{0.48\textwidth}
\centering
\includegraphics[width=\textwidth]{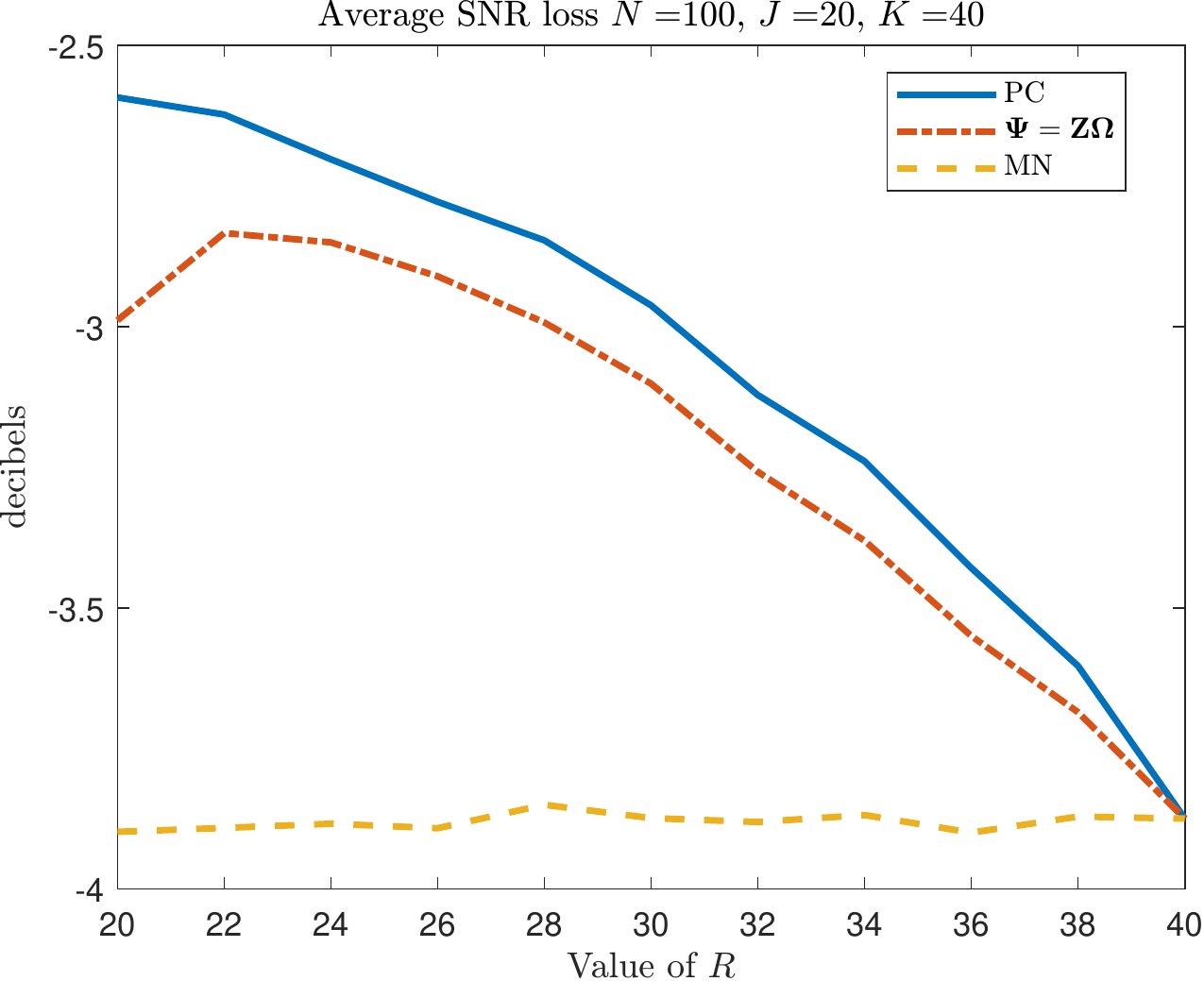}
\caption{$J=20$, $K=40$}
\end{subfigure}
\caption{Average SNR loss versus $R$. $N=100$ and $K=2J$.}
\label{fig:mean_snrloss_vs_R}	
\end{figure}

We then study the influence of $R$ in Figure \ref{fig:mean_snrloss_vs_R} where we also plot the SNR loss of  the adaptive filter given in \eqref{w_MN} which is denoted as MN. Again we observe that it is of interest to choose $R$ slightly larger than $J$, and that when $J/N$ increases the method based on randomized projections comes very close to the PC method. As for the MN method, its loss is rather small for $J=10$ but increases when $J=20$.

\begin{figure}[htb]
\centering
\begin{subfigure}[b]{0.48\textwidth}
\centering
\includegraphics[width=\textwidth]{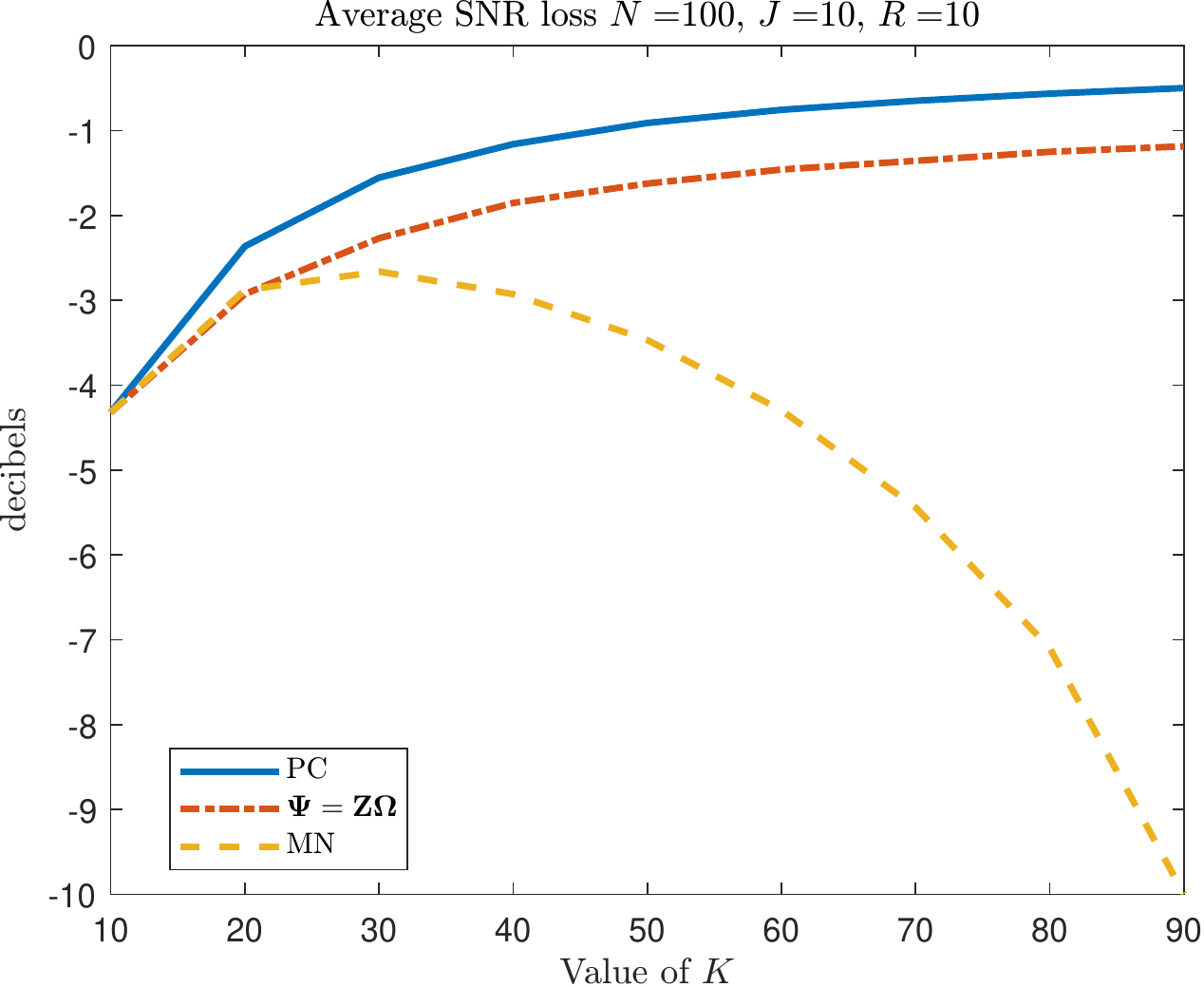}
\caption{$J=10$, $R=10$}
\end{subfigure}
\hfill
\begin{subfigure}[b]{0.48\textwidth}
\centering
\includegraphics[width=\textwidth]{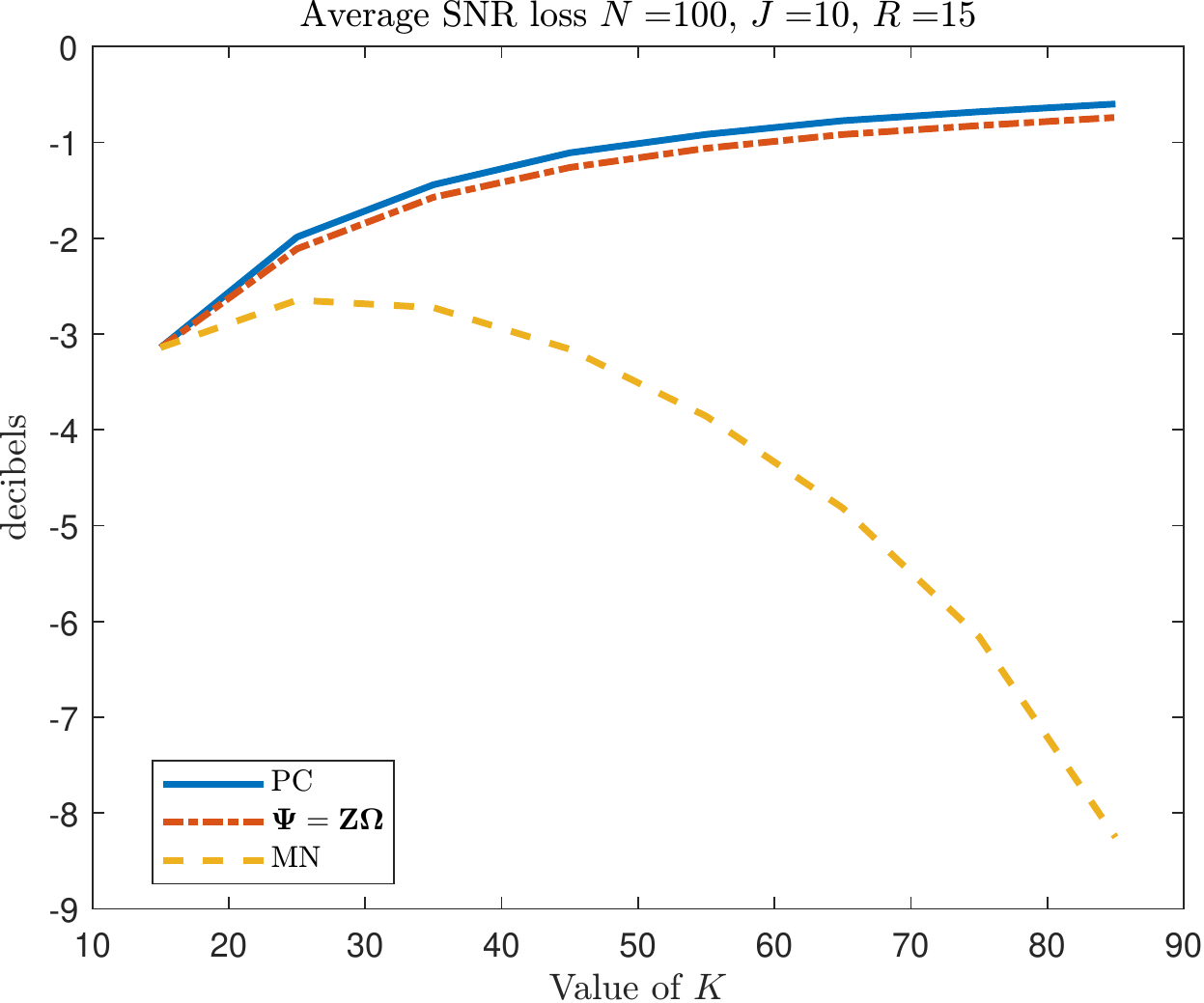}
\caption{$J=10$, $R=15$}
\end{subfigure}
\caption{Average SNR loss versus $K$. $N=100$ and $J=10$.}
\label{fig:mean_snrloss_vs_K}	
\end{figure}

Finally the influence of $K$ is studied in Figure \ref{fig:mean_snrloss_vs_K} where $J=10$ and $R=10$ or $R=15$. One can observe that the MN method performs very well when $K$ is small, typically $K \gtrsim J$ but rapidly degrades with $K$ increasing. The difference between the PC method and the method based on $\mPsi=\Z\mOmega$ tends to increase slightly with $K$.

\section{Conclusions}
We presented an alternative to the principal component method where the interference subspace is estimated using randomized projections, following ideas proposed recently to achieve low-rank matrix approximations at reduced cost. The basic idea is to estimate the principal subspace of the interference in the auxiliary channels $\Z$ by $\Z\mOmega$ where $\mOmega$ is random. We showed that the method depends weakly on the choice of $\mOmega$, provides an SNR very close to that of the PC method provided that the number of columns in $\mOmega$ is chosen slightly above the rank of the low-rank component of the interference. The new method thus provides a computationally interesting way to design partially adaptive filters when the number of training samples is low.

\end{document}